\documentclass[12pt]{iopart}
\usepackage{iopams}

\usepackage[russian,english]{babel}
\usepackage{amsfonts}
\usepackage{amssymb}
\usepackage{amsthm}
\usepackage{amscd}
\usepackage{epsfig}

\usepackage[latin1]{inputenc}
\usepackage{a4wide}

\def\be{\begin{equation}}
\def\ee{\end{equation}}
\def\bea{\begin{eqnarray}}
\def\eea{\end{eqnarray}}

\def\a{\alpha}
\def\b{\beta}

\def\k{\kappa}
\def\s{\sigma}

\newcommand{\Sc}{Schr\"odinger }

\newcommand{\rf}[1]{(\ref{#1})}
\newcommand{\eqref}[1]{(\ref{#1})}
\newcommand{\nn}{\nonumber}

\begin{document}

\title[Non-conservative
multichannel
SUSY partners of the zero potential]{Spectral properties of non-conservative
multichannel
SUSY partners of the zero potential}

\author{Andrey M Pupasov\footnote{Boursier de l'ULB}$^{1, 2}$, Boris F Samsonov$^1$ and Jean-Marc Sparenberg$^2$}

\address{$^1$ Physics Department, Tomsk State University, 36 Lenin Avenue,
634050 Tomsk, Russia}

\address{$^2$ Physique Quantique, C.P.\ 229, Universit\'e Libre de Bruxelles,
B 1050 Bruxelles, Belgium}

\eads{\mailto{pupasov@phys.tsu.ru}, \mailto{samsonov@phys.tsu.ru}, \mailto{jmspar@ulb.ac.be}}

\begin{abstract}
Spectral properties of a coupled $N \times N$ potential model obtained with
  the help of a single non-conservative supersymmetric (SUSY) transformation
starting from a system of $N$ radial Schr\"odinger equations with the
zero potential and finite threshold differences between the channels are studied.
The structure of the system of polynomial equations which determine the zeros
of the Jost-matrix determinant is analyzed.
In particular, we show that the Jost-matrix determinant has $N2^{N-1}$ zeros
which may all correspond to virtual states.
The number of bound states satisfies $0\leq n_b\leq N$.
The maximal number of resonances is $n_r=(N-1)2^{N-2}$.
A perturbation technique for a small coupling
approximation is developed.
A detailed study of the inverse spectral problem is given for the $2\times 2$ case.
\end{abstract}



\section{Introduction}

Almost all low-energy collisions
of microparticles with an internal structure (i.e., atom-atom, nucleus-nucleus etc)
include
inelastic processes such as excitations of internal degrees of
freedom of colliding particles
or processes with rearrangements of their constituent
parts.
These processes can be described by a matrix
(more precisely multichannel)
\Sc equation with a
local matrix potential \cite{taylor:72,newton:82}.
One may be
interested in both direct and inverse scattering problems for
this equation.
The method of SUSY transformations is known as a
powerful tool for solving both types of problems for a single-channel
\Sc equation \cite{sukumar:85}.
Nowadays, the first attempt to
generalize the method for a coupled-channel \Sc equation
with different thresholds is given in
\cite{sparenberg:06,samsonov:07}. This attempt is based on a non-conservative
SUSY transformation (contrary to \cite{amado:88a,amado:88b}),
i.e. a SUSY transformation that does not preserve a boundary
behavior of solutions. The main advantage of such transformations
is a possibility to obtain multichannel potentials with a non-trivial
coupling starting from the zero potential.

The present work is aimed at the investigation of spectral
properties of these SUSY potentials. Our approach is based on an analysis of the
Jost matrix.  In the non-relativistic scattering theory the Jost
matrix plays a fundamental role similar to the scattering matrix.
The zeros of
the Jost-matrix determinant define positions of the bound/virtual states and resonances
\cite{taylor:72,newton:82}.
Therefore, studying  the zeros of the Jost-matrix determinant
allows one to analyze the spectrum of the model.
A closed analytical
expression of the Jost matrix, as well as potential, resulting from a
non-conservative SUSY transformation
of the zero potential
 is obtained in
\cite{sparenberg:06}.
The analysis of spectral properties for such
potentials was not presented up to now despite the fact that the
Jost matrix is well known \cite{cox:64}.
This may be explained by the fact that
the spectrum of the potential
after a
non-conservative SUSY transformation changes essentially
and to find these changes one has to find all the zeros of
the Jost-matrix determinant.
 More precisely, no
one spectral point of the initial
Hamiltonian
belongs to the spectrum of the transformed Hamiltonian.
 As a result, a supersymmetry algebra,
which is always present in the case of conservative SUSY
transformations, cannot actually be constructed here and the word
'SUSY transformation' is only a formal heritage from the
previous conservative case \cite{amado:88a,amado:88b}.

The principal point of this paper is
to show
that the qualitative behavior
of the spectrum of (non-conservative) SUSY partners of the
vanishing multichannel potential with threshold differences may be
studied for an arbitrary number of channels, $N$.
We think this
is a very strong result, since even for the case $N=2$ the full analysis of
the spectrum is a very complicated problem \cite{cox:64,pupasov:07,pupasov:08}.
The main reason for
this is an extremely rapid growth of the order of an algebraic
equation defining the spectrum with the growth of the number
of channels.

The paper is organized as follows. We start with
 preliminaries, where we give basic definitions and equations.
  Section \ref{sec:NBS} is devoted to the analysis of
the number of bound states resulting from a non-conservative SUSY
transformation of the zero potential as a function of the parameters
defining the transformation.
This analysis is based on the
study of the properties of the eigenvalues of the Jost matrix.
Following similar lines we analyze the possible number
 of virtual states in section
\ref{sec:VS}.
Once the bound and virtual states are analyzed
we can formulate conditions under which resonances may appear; this is made in section \ref{sec:R}.
The behavior of the Jost-matrix determinant zeros is studied in section \ref{sec:WC} in
the approximation of a weak coupling between channels. In section \ref{sec:Cox} we
deal with the particular two-channel case. In this case we express parameters of the potential
in terms of zeros of the Jost-matrix determinant, i.e. solve an inverse spectral problem.
The main results are summarized in the conclusion.

\section{\label{sec:Pr}Preliminaries}

Let us first summarize the notations used below for coupled-channel scattering theory
\cite{taylor:72,newton:82,vidal:92a}.
We consider a system of coupled radial Schr\"odinger equations for the $s$-waves
that in reduced units reads
\begin{equation}\label{schr}
H\psi(k,r)=K^2\psi(k,r),\quad r\in(0,\infty)
\end{equation}
with
\begin{equation}
H=-\mathbf{1}\frac{d^2}{d r^2}+V(r),
\end{equation}
where $r$ is the radial coordinate, $V(r)$ is an $N\times N$ real symmetric matrix,
$\mathbf{1}$ is the unit matrix,
and $\psi$ may be either a matrix-valued or a vector-valued solution.
By $k$ we denote a point in the space ${\mathbb C}^N$,
$k=\left\{k_1,\ldots,k_N\right\}$,
$k_i\in \mathbb C$.
A diagonal matrix with non-vanishing entries $k_i$ is written as
$K=\mbox{diag}(k)=\mbox{diag}(k_1,\ldots,k_N)$.
The complex wave numbers $k_i$ are related to the center-of-mass energy $E$
and the channel thresholds $\Delta_1,\dots, \Delta_N$,
which are supposed to be different from each other,
$\Delta_{i(\ne j)}\ne\Delta_j$,
by
\begin{equation}\label{thrE}
k_j^2=E-\Delta_j\,,\qquad\Delta_1=0\,.
\end{equation}
We assume here that $\Delta_1=0$ and the different channels have
equal reduced masses, a case to which the general situation can
always be formally reduced \cite{newton:82}.

Let us recall basic definitions from SUSY quantum mechanics \cite{sukumar:85,sparenberg:06,samsonov:07,amado:88a,amado:88b}.
It is known that the solutions of the initial \Sc equation \eqref{schr} may be
mapped into the solutions of the transformed equation
with help of the differential-matrix operator
\begin{equation}\label{psit}
\tilde{\psi}(k,r)=L\psi(k,r)=\left[-\mathbf{1}\frac{d}{dr}+U(r)\right]\psi(k,r)\,.
\end{equation}
The transformed \Sc equation has form \eqref{schr} with a new potential
\begin{equation}\label{Vt}
\tilde{V}(r)=V(r)-2 U'(r)\,.
\end{equation}
Matrix $U$ is called superpotential
\begin{equation}\label{U}
U(r)=\eta'(r) \eta^{-1}(r)\,,
\end{equation}
and expressed in terms of a matrix solution $\eta$ of the initial
\Sc equation
\begin{equation}\label{shrtr}
H \eta(r) = -{\cal K}^2 \eta(r)\,,
\end{equation}
where ${\cal
K}=\mbox{diag}(\kappa)=\mbox{diag}(\kappa_1,\dots,\kappa_N)$ is a
diagonal matrix called the factorization wave number, which
corresponds to an energy $\cal E$ lying below all thresholds,
called the factorization energy. The entries of $\cal K$, thus,
satisfy ${\cal E}=-\kappa_i^2+\Delta_i$; by convention, we choose
them positive: $\kappa_i>0$. Solution $\eta$ is called the factorization solution.

In the case of the zero potential $V=0$, $\eta$
contains only exponentials
\begin{equation}
\eta(r)=\cosh({\cal K} r) + {\cal K}^{-1} \sinh({\cal K} r) U_0\,.
\end{equation}
The symmetric matrix $U_0$ is the superpotential at $r=0$, which can be chosen
arbitrary.
It is convenient to introduce special notations $\alpha_j$ for the diagonal and
$\beta_{jl}$ for the off-diagonal entries of $U_0$.

The Jost matrix of a (non-conservative) SUSY partner of the
$N$-channel zero potential reads \cite{sparenberg:06}
\begin{equation}\label{FtCox}
F(k)=({\cal K}-i K)^{-1}(U_0-i K)\,,
\end{equation}
which is also the Jost matrix obtained in \cite{cox:64}.

The necessary and sufficient condition
 on the parameters (factorization energy ${\cal E}$ and superpotential at the origin $U_0$)
 to get a potential without singularity
 at finite distances is obtained in \cite{pupasov:07,pupasov:08}.
 This condition is the positive definiteness of matrix
 ${\cal K}+U_0$:
\begin{equation}\label{posU}
{\cal K}+U_0>0\,,
\end{equation}
which puts some upper limit on the factorization energy ${\cal E}$
at fixed $U_0$.

Zeros of the Jost-matrix determinant define positions of the
bound/virtual states and the resonances. Thus, to find these positions
we have to solve the following equation
\begin{equation}\label{jdz}
{\rm det}F(k)=0\,,
\end{equation}
taking into account the threshold conditions \eqref{thrE}.
According to \eqref{FtCox}, the roots of equation \eqref{jdz}
are defined by the roots of
\begin{equation}\label{bz}
{\rm det}B(k)=0\,,\qquad \kappa_j-ik_j\neq 0\,,\qquad j=1,\ldots, N\,,
\end{equation}
where
\begin{equation}
\label{bm}
B(k)=U_0-i K\,.
\end{equation}
In what follows we concentrate on the analysis of the zeros of ${\rm det}B$ only
keeping in mind that some of them may be cancelled in ${\rm det}F$ if
$k_j=-i\k_j$.
Our starting point is thus a system of algebraic equations
\eqref{bz} and \eqref{thrE}
 which reads, with certain coefficients $a_i^j$,
\begin{eqnarray}
(-i)^Nk_1k_2\ldots k_N+\sum\limits_{j=1}^N a_{N-1}^j\prod\limits_{l=1,\,l\neq j}^N k_l+
\ldots+\sum\limits_{j=1}^Na_1^jk_j+a_0  =  0\,,\label{algeq}\\
k_j^2-k_1^2+\Delta_j  =  0\,.\label{algeqa}
\end{eqnarray}

First we show that system \eqref{algeq}, \eqref{algeqa}
 can be reduced to an algebraic equation
of the  $N 2^{N-1}$ degree with respect to one momentum, say $k_1$,
only.
Indeed, any momentum enters equation  \eqref{algeq} only linearly.
Therefore it can be rewritten in the form
\be\label{kN}
k_NP_{1}(k_1,\ldots,k_{N-1})=Q_{1}(k_1,\ldots,k_{N-1})\,,
\ee
where $P_{1}(k_1,\ldots,k_{N-1})$ and
$Q_{1}(k_1,\ldots,k_{N-1})$ are polynomials of
the first degree in each of the variables  $k_1,\ldots,k_{N-1}$.
It is important to note that given all momenta
$k_1,\ldots,k_{N-1}$ this equation defines $k_N$ in a
unique way if $P_{1}$ does not vanish. On the other hand we can square the left- and
 right-hand sides of \eqref{kN} thus obtaining an equation
 where $k_N$ enters only in the second degree and
 polynomials $P_{1}^2$ and $Q_{1}^2$ are polynomials of
 the second degree with respect to their variables.
 But in the equation thus obtained
using threshold condition \eqref{algeqa}
 we can replace all second powers of the variables $k_j$,
 $j=2,\ldots,N$
 by  $k_1^2-\Delta_j$,
 which makes disappear both
 variable $k_N$
and the second power of $k_j$, $j=2,\ldots,N-1$  from the resulting
equation and raises the
 power of $k_1$ till $2N$.
 We thus see that after these
 manipulations variable $k_{N-1}$ enters in the resulting
 equation only in the first degree and the equation can be
 rewritten in form \rf{kN}
\be\label{kN-1}
k_{N-1}P_{2}(k_1,\ldots,k_{N-2})=Q_{2}(k_1,\ldots,k_{N-2})\,,
\ee
where $P_{2}(k_1,\ldots,k_{N-2})$ and
$Q_{2}(k_1,\ldots,k_{N-2})$ are polynomials of
the first degree in each of the variables $k_2,\ldots,k_{N-2}$.
From \eqref{kN-1}, given $k_1,\ldots,k_{N-2}$, not a zero of $P_{N-2}$, we obtain $k_{N-1}$ in a unique way.
We note that the system \eqref{kN-1}, \eqref{kN} and \eqref{algeqa} where
from \eqref{algeqa} the last equation $k_N^2-k_1^2+\Delta_N=0$
should be excluded, is equivalent to the original system \eqref{algeq},
 \eqref{algeqa}.

It is clear that we can repeat the above process $N-3$
times more to get an equation
\be\label{k2}
k_2P_{N-1}(k_1)=Q_{N-1}(k_1)
\ee
and finally
\be\label{k1}
P_N(k_1)=0
\ee
with $P_N$ of order $N2^{N-1}$. Note, that the subscript in $P_k$ and $Q_k$
indicates nothing but the step in this procedure.
It is evident that any $k_1$ which (together with
$k_2,\ldots,k_N$) solves the system \rf{algeq}, \rf{algeqa}
is a root of \rf{k1}.
The converse is also true. Indeed, given a root $k_1$
of \rf{k1}, but not a root of $P_{N-1}$, we find from \rf{k2} a unique $k_2$. Once we
know $k_1$ and $k_2$ we find $k_3$ from equation previous
to \rf{k2} and so on till $k_N$ which is found from
\rf{kN}.
It is also clear that in this way we can get $N2^{N-1}$
number of sets $k_1,\ldots,k_N$
(some of them may coincide)
each of which solves the
system \rf{algeq}, \rf{algeqa} so that the same number
 $N2^{N-1}$ is the number of possible solutions of this
 system and the system \eqref{k1}, \eqref{k2}, \ldots, \eqref{kN} is equivalent
 to the initial system \eqref{algeq}, \eqref{algeqa}.

\section{\label{sec:NBS} Number of bound states}

In the following, except for sections \ref{sec:WC} and \ref{sec:Cox},
we will consider all quantities as functions  of the momentum $k_1$.
Other momenta are expressed in terms of $k_1$ from the threshold conditions \eqref{thrE}.
Since in this section we are interested in the number of bound states
we will consider only the negative energy semi-axis $E\in(-\infty,0)$.
It happens to be useful to change variables $k_j$ in favor
of $\bar k_j$ as $k_j=i\bar{k}_j$ and rewrite the threshold
conditions \rf{algeqa} accordingly
\begin{equation}\label{kj}
\bar{k}_j=\sqrt{\bar{k}_1^2+\Delta_j}\,,
\end{equation}
where
 we have chosen only the positive value of the square root
 since
in this section we analyze only the point spectrum of $H$,
which restricts all momenta $k_j$ to be purely imaginary
with a positive imaginary part so that $\bar k_j=|k_j|$.

From  \eqref{FtCox} it is clear that all the zeros of $\mathrm{det}{F}$
are at the same time the zeros of the determinant of matrix
$B$ \rf{bm} and vice versa.
This follows from \eqref{bz} and the positive definiteness
of matrix ${\cal K}-iK$ in the momenta region we consider so that
neither of the roots of $\mathrm{det}\ {B}$ solves the
equation $\mathrm{det}({\cal K}-iK)=0$.

Since $\mathrm{det}\ {B}=\prod_{j=1}^N\lambda_j$
where $\lambda_j$ are the eigenvalues of $B$,
\begin{equation}\label{es}
B(\bar k_1)\,{x}_j(\bar k_1)=\lambda_j(\bar k_1)\,{x}_j(\bar k_1)\,,
\end{equation}
the equation $\mathrm{det}{B}(\bar k_1)=0$ is equivalent to
$\lambda_j(\bar k_1)=0$, $j=1,\ldots,N$.
Matrix $B$ is symmetric with real entries
in the momenta region we consider, $B=U_0+\bar{K}=B^T$, which implies the reality of both
$\lambda_j(\bar k_1)$ and $x_j(\bar k_1)$. Here we introduced a diagonal matrix $\bar{K}=|K|=\mbox{diag}(\bar
k_1,\ldots,\bar k_N)$.

Another property of $\lambda_j(\bar k_1)$ important for the analysis
 is their monotony as functions of $\bar k_1$
that we prove below.

For a fixed $\bar{K}$ let us consider
a deviation of
$\lambda_j(\bar{k}_1)$ for a small increment of argument $\bar k_1$, i.e.
$\lambda_j(\bar k_1+\delta \bar k_1)=\lambda_j(\bar k_1)+\delta\lambda_j(\bar k_1)$
assuming
$\delta\bar K=\mbox{diag}(\delta\bar k_1,\ldots,\delta\bar k_N)$
real, positive definite (since $\delta\bar k_j>0$, $\forall j$)
and infinitesimal.
From \eqref{es} one gets
\begin{equation}\label{esp}
B(\bar{k}_1+\delta \bar{k}_1){x}_j(\bar{k}_1+\delta \bar{k}_1)=
\lambda_j(\bar{k}_1+\delta \bar{k}_1){x}_j(\bar{k}_1+\delta \bar{k}_1)\,.
\end{equation}
Here according to \eqref{bm}
$B(\bar{k}_1+\delta \bar{k}_1)=U_0+\bar{K}+\delta \bar{K}$
and the increment of $B(\bar k_1)$ is just
 $\delta B=\delta \bar{K}$
which plays the role of a small perturbation of $B(\bar k_1)$.
Therefore
we may calculate the shifting of the eigenvalues
produced by such a perturbation
 using a (Rayleigh-\Sc$\!\!$) perturbation theory.
Thus, for a non-degenerate eigenvalue $\lambda_j$
 the first order correction reads
\begin{equation}
\delta \lambda_j=\langle x_j|\delta B|x_j\rangle>0
\end{equation}
where the inequality follows from the positive definiteness
of $\delta B=\delta\bar K$,
which in turn implies monotony of the eigenvalues as
functions of the momenta $\bar k_1$.
For a degenerate eigenvalue corrections are obtained by
diagonalizing the same perturbation operator $\delta B$
restricted to a linear span of unperturbed eigenvectors
corresponding to a given eigenvalue, which still leads to positive
corrections because of positive definiteness of $\delta B$.

From here it follows that any
eigenvalue $\lambda_j(\bar k_1)$ may vanish i.e. change its sign, only once.
Moreover, $\lambda_j\rightarrow\bar{k}_j>0$ as $\bar k_1\rightarrow\infty$.
Hence, the number of
negative eigenvalues of $B$ at $\bar k_1=0$,
 i.e.
at the energy of the lowest threshold,
is just the number of bound states.
Thus, to count the number of the bound states, $n_b$,
one has to consider the
eigenvalues $\lambda_j(\bar k_1)$,
 $j=1,\dots,N$
 of
matrix $B(\bar k_1)$ at $\bar k_1=0$,
\begin{equation}
B(0) \equiv U_0-i\ \mathrm{diag}\ (i\sqrt{\Delta_j})=
 U_0 + \mathrm{diag}(\sqrt{\Delta_j})
\end{equation}
so that
\begin{equation}\label{nb}
n_b=\frac{1}{2}(N-\Lambda)\,,\qquad
\Lambda=\sum_{j=1}^N\Lambda_j\,,
\qquad \Lambda_j=\frac{\lambda_j(0)}{|\lambda_j(0)|}\,.
\end{equation}
To clarify this formula we notice that
in the absence of bound states
all $\Lambda_j=1$, $\Lambda=N$ so that $n_b=0$.
Every bound state is responsible for the change of the sign
 of only one eigenvalue from positive to negative thus
 raising $-\Lambda$ by $2$ units, i.e.
 $-\Lambda \rightarrow
 - \Lambda+2$ with $n_b\rightarrow n_b+1$.
This justifies the factor $1/2$ in \eqref{nb}.
\begin{figure}
\begin{center}
\epsfig{file=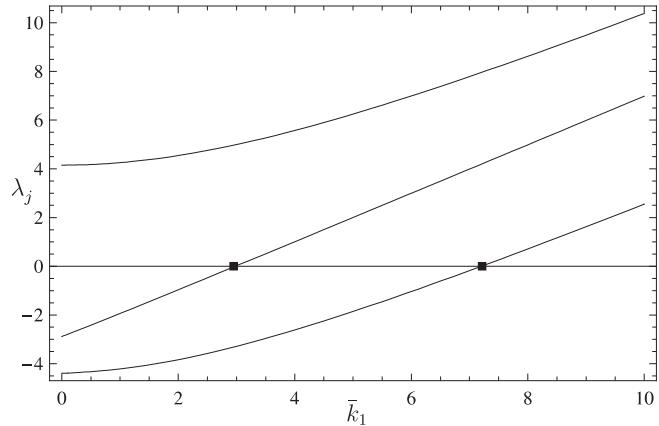, width=8.6cm}
\caption{\small Typical behavior of $B$-matrix eigenvalues, $N=3$.
The case of two bound states with energies $E_1=-51.8611$ and $E_2=-8.8852$ is presented.
 The black squares show positions of these bound states. The corresponding parameters
 are $\alpha_1=-3$, $\alpha_2=-8$, $\alpha_3=-1$,
 $\beta_{12}=0.5$, $\beta_{13}=0.4$, $\beta_{23}=1$, $\Delta_2=15$, $\Delta_3=25$.
\label{figEV}}
\end{center}
\end{figure}

Summarizing, we see that the number of bound states is bounded
by $0\leq n_b\leq N$. Figure \ref{figEV} shows the eigenvalues of matrix $B$ as functions
of $\bar{k}_1$ for the case $N=3$. Two eigenvalues cross the axis which corresponds to the
case of $n_b=2$. The last comment in this section is devoted to
equation \eqref{posU}. Now it can be seen that the factorization
energy should be chosen lower than the ground-state energy for the transformed potential, ${\cal E}<E_g$,
if any.

\section{\label{sec:VS} Number of virtual states}

According to the definition of a virtual state \cite{taylor:72,newton:82}
in this section we will need to consider the channel wave numbers
$k_j$
lying both in the positive and the negative imaginary semi-axes
of the corresponding momenta complex planes
and consider
the full imaginary axis for $k_1$, i.e.
 $\bar{k}_1\in(-\infty,\infty)$.
The other momenta, $k_2,\ldots,k_N$, belong to
either the positive or to negative
parts of their imaginary axes in agreement with the
threshold conditions
\begin{equation}\label{kjV}
\bar{k}_j=\sigma_j\sqrt{\bar{k}_1^2+\Delta_j}\,,\qquad\sigma_j=\pm\,,\qquad
j=2,\ldots,N\,.
\end{equation}

Since in \eqref{kjV} all combinations of signs are now possible
it is convenient to introduce special notations for these
combinations.
Denote $\sigma=(+,\pm,\ldots,\pm)$ a string of
$N$ signs with $\sigma_j$ being its $j$-th entry, which
corresponds to the sign in \rf{kjV} for the $j$-th momentum for
$j>1$. The first symbol "$+$" in $\sigma$ indicates that all momenta
$\bar k_j$ are expressed in terms of $\bar k_1$.
Let $n_{+}(\s)+1$ and $n_{-}(\s)$ be the numbers of "$+$" and "$-$"
signs in this string.
We notice the following evident combinatoric properties of
$n_{-}(\sigma)$ and $n_{+}(\sigma)$.
First, $n_{+}(\sigma)+n_{-}(\sigma)+1=N$ which implies
\begin{equation}\label{summ}
\sum\limits_{\sigma}\left[n_+(\sigma)+n_-(\sigma)+1\right]=N2^{N-1}\,.
\end{equation}
Here and in what follows the summation over $\s$ includes
all $2^{N-1}$ possible sign combinations.
Next,
a symmetry between "$+$" and "$-$" leads to the following relation
\begin{equation}\label{sympm}
\sum\limits_{\sigma}n_-(\sigma)=\sum\limits_{\sigma}n_+(\sigma)=(N-1)2^{N-2}\,.
\end{equation}

 According to \rf{bm} every sign combination leads to its
own $B$ matrix defined by the corresponding $K$ matrix so
that both $K$ and $B$ should carry an additional
information about this combination.
Therefore
\begin{equation}
\label{vm}
B^{\sigma}=U_0+\bar{K}^{\sigma}\,,\qquad \bar{K}^{\sigma}=
{\rm diag} (\bar{k}_1,\sigma_2\bar{k}_2,\ldots,\sigma_N\bar{k}_N)
\end{equation}
and we denote $\lambda_j^\sigma(\bar{k}_1)$,
$j=1,\ldots,N$  the eigenvalues of $B^{\sigma}$.

In order to find the zeros of the Jost-matrix determinant
corresponding to the
virtual states we should find the purely real solutions of the equations
$\lambda_j^\sigma(\bar{k}_1)=0$, $j=1,\ldots,N$ for
all $2^{N-1}$ matrices $B^{\sigma}$.
Although the $\bar{k}_j$'s are real, but bearing in mind our
replacement $k_j=i\bar{k}_j$,
throughout the text
 we call these zeros purely imaginary.
 Finally we note that
  since matrix ${\cal K}-iK$ in \rf{FtCox}
  is not positive definite for an arbitrary
$\sigma$ anymore,
in some particular cases
 some of the zeros of $B$ may be cancelled
by the zeros of $\mbox{det}({\cal K}-iK)$
  and will not correspond to virtual states.
  Nevertheless, omitting these particular cases, we will
  concentrate on an analysis of the zeros of $\mbox{det}B$
  only.

\begin{figure}
\begin{center}
\begin{minipage}{15cm}
\epsfig{file=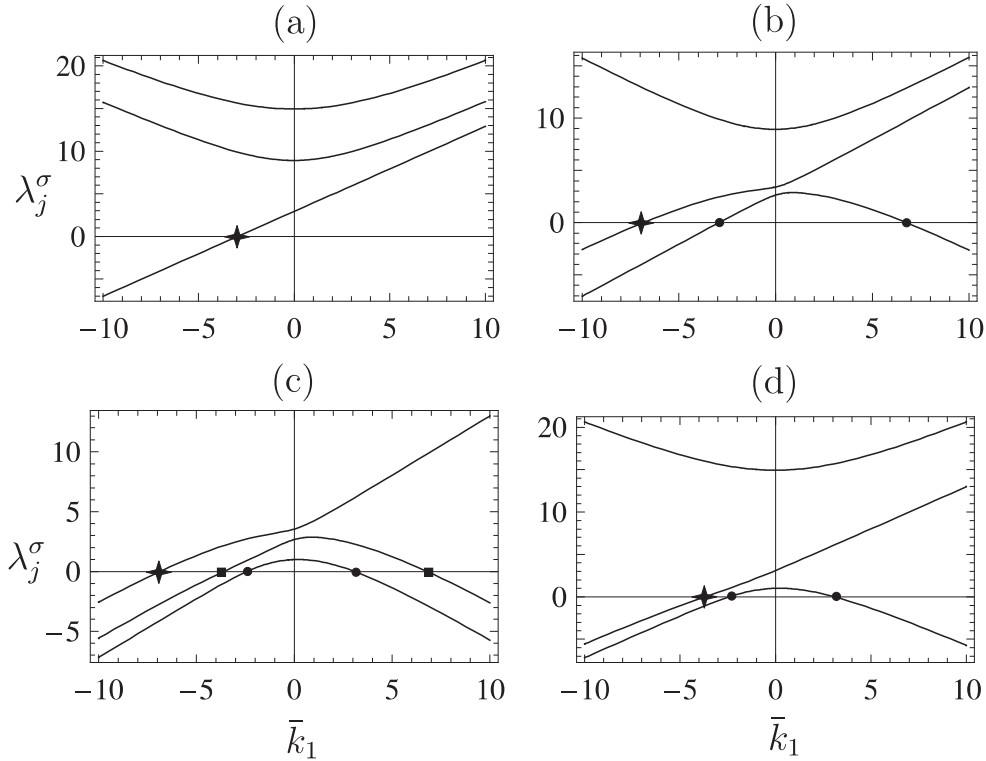, width=13cm} \caption{\small Typical behavior
of the eigenvalues $\lambda_j^\sigma(\bar{k}_1)$, $N=3$, is shown. Each
plane corresponds to a particular choice of string $\sigma$:
(a) $\sigma=(+++)$, (b) $\sigma=(++-)$, (c) $\sigma=(+--)$, (d) $\sigma=(+-+)$.
Stars, squares and circles correspond
to the virtual states. Virtual states are denoted by the identical symbol
if they belong to the same eigenvalue $\lambda_j^\sigma(\bar{k}_1)$.
The corresponding
parameters are
$\alpha_1=3$, $\alpha_2=5$, $\alpha_3=9$, $\beta_{12}=0.5$, $\beta_{13}=0.4$,
 $\beta_{23}=0.2$, $\Delta_2=15$, $\Delta_3=35$.
\label{fig4Planes}}
\end{minipage}
\end{center}
\end{figure}

Eigenvalues $\lambda_j^\sigma(\bar{k}_1)$ are monotonous functions of $\bar{k}_1$
in
two cases only: (i) $\sigma=(+,+,\ldots,+)$ and $\bar{k}_1>0$;
(ii) $\sigma=(+,-,\ldots,-)$ and $\bar{k}_1<0$.
In general, an eigenvalue $\lambda_j^\sigma(\bar k_1)$
may have minima/maxima for $\bar{k}_1\lessgtr0$  which may lead to two or even more
 roots in equation $\lambda_j^\sigma(\bar k_1)=0$.
 We illustrate this behavior for $N=3$ in
figure~\ref{fig4Planes}.
The monotonous lines in the right/left part of figure~\ref{fig4Planes}(a)/(c).
correspond to case (i)/(ii).
The position of the zeros of the eigenvalues is shown
by stars, squares and circles.
It is clearly seen that the total number of the roots of
all equations $\lambda_j^\sigma(\bar k_1)=0$ is
$(N2^{N-1})|_{N=3}=12$ which all correspond to virtual states.

A change of parameters may result in shifting the position of the virtual states only
without changing the number of zeros
(i.e. virtual states).
For instance, in
the simplest case  we may shift all diagonal entries
of $U_0$ by a number $\lambda_0$,
 $U_0\to U_0+\lambda_0I$,
 thus shifting all eigenvalues of
$B$ by the same number,
${\lambda}_j^\sigma(\bar{k}_1)\to
\lambda_j^\sigma(\bar{k}_1)+\lambda_0$.

Let us consider a specific eigenvalue defined by a string
$\sigma_0$,
 with a local maximum at $\bar k_1=\bar k_{1,max}$,
$\lambda_j^{\sigma_0}(\bar{k}_{1,max})=\lambda_{j,max}$.
One can always shift all the eigenvalues by the value
$\lambda_{j,max}$ such that the curve
$\lambda_j^{\sigma_0}(\bar{k}_1)$ touches the $\bar k_1$ axis
at the point $\bar k_1=\bar{k}_{1,max}$
meaning that $\bar{k}_{1,max}$ not only becomes a root of
the equation $\lambda_j^{\sigma_0}(\bar{k}_1)=0$ but this
root is multiple (of multiplicity 2) and by a small additional change of
other parameters it can be split into two simple but complex roots.
This is just in this way two virtual states collapse
producing a resonance; a subject which deserves a special
discussion (see the next section). Pairs of virtual states which may collapse are shown in figure~\ref{fig4Planes} by
squares and circles.

It is not difficult to convince oneself
 that for any given $\b_{jl}$ the situation when all the zeros
of the Jost-matrix determinant are purely
imaginary may be realized by a proper choice of $\alpha_j$.
To see that
let us consider the asymptotic behavior of $\lambda_j^\sigma$ for
$|\bar{k}_1|\rightarrow\infty$, when all off-diagonal
entries of $B$ become negligibly small,
\begin{eqnarray}\label{asymb}
\lambda_1^\sigma & \simeq & \bar{k}_1+\alpha_1\,,\\
\lambda_j^\sigma & \simeq & \sigma_j\sqrt{\bar{k}_1^2+\Delta_j}+\alpha_j=
\sigma_j\left(|\bar{k}_1|+\frac{\Delta_j}{2\bar{k}_1}+\ldots\right)+\alpha_j\,.\\
|\bar{k}_1| & \rightarrow & \infty\,.
\end{eqnarray}
Numbers $n_+(\s)$ and $n_-(\s)$
determine the corresponding numbers of increasing and decreasing
eigenvalues at positive infinity.
The eigenvalue $\lambda_1^\sigma$ increases both at negative
and positive infinity.
Now if we choose all $\a_j$ sufficiently large in absolute
values and negative we can always guaranty the location of
a root of the equation $\lambda_1^\sigma(\bar k_1)=0$ near
the point $\bar k_1=\a_1$
and at the same time the location of two roots of the
equation $\lambda_j^\sigma(\bar k_1)=0$ with corresponding
$\sigma_j=+$ near the points
 $\bar k_1=\pm\a_j$.
Thus, for each $\sigma$ we can obtain $2n_+(\sigma)+1$ zeros.
The total number $n_v$ of these zeros
may be calculated by formulas \eqref{summ} and \eqref{sympm}
\begin{equation}
n_v=\sum\limits_{\sigma}[2n_+(\sigma)+1]=N2^{N-1}\,,
\end{equation}
which coincides with the total number of all possible roots
of the system \rf{algeq}, \rf{algeqa}
and is just the maximal possible number of virtual
states.
Hence, in this case all the roots are purely imaginary.
In the next section we consider the case when some of the zeros
may merge, become complex and produce resonances.

\section{\label{sec:R} Number of resonances}

For simplicity independently on whether or not it can be seen
in a scattering we call any pair of complex zeros $k=\pm k_r+ik_i$
of the Jost-matrix determinant a resonance keeping in mind
that to be really visible in a scattering a resonance behavior
of the corresponding cross-section
should be narrow and sharp enough.

Conservation of the number of zeros of an $n$-th order
algebraic equation under a variation of parameters included in
its coefficients, which keeps unchanged its order
(in our case this is equation \rf{k1} obtained from the
system \rf{algeq}, \rf{algeqa}) applied to our case
leads to the following relation
 $n_b+n_v+2n_r=N2^{N-1}$, where $n_b$, $n_v$ and $n_r$ are number of
bound states, virtual states and resonances respectively.
The aim of this section is to establish the maximal number of
possible resonances
accepted by a non-conservative SUSY-partner of the vanishing potential.

Evidently, the maximal number of
resonances corresponds to the minimal number of bound
$n_b$ and virtual $n_v$ states.
These numbers would both become zero if no one of
the $B$ matrix eigenvalues  intersected the $\bar k_1$
axis. But as it was noticed in the previous section there
always exists an eigenvalue
$\lambda_1^\s$
with the asymptotic behavior
given in \rf{asymb}, i.e. ranging from $-\infty$ to
$+\infty$ and, hence, it intersects $\bar k_1$ axis always
and for all possible values of $\s$.
We thus see  that the
minimal number of real zeros that all eigenvalues may take
is achieved if all eigenvalues
$\lambda_j^\sigma(\bar{k}_1)$, $j>1$ are nodeless and
curves
$\lambda_1^\sigma(\bar{k}_1)$ intersect $\bar k_1$ axis
only once for every given sign combination $\s$.
To
realize this case, we should choose parameters included in
$U_0$ in a such way
that the global minimum $\lambda_{j,min}^\sigma$ of every eigenvalue
$\lambda_j^\sigma(\bar{k}_1)$
 with $\sigma_j=+$
(they tend to $+\infty$ when $|\bar k_1|\to\infty$)
 be positive $\lambda_{j,min}^\sigma>0$
 and, respectively,
global maximum $\lambda_{j,max}^\sigma$
of every eigenvalue
$\lambda_j^\sigma(\bar{k}_1)$
 with $\sigma_j=-$
(they tend to $-\infty$ when $|\bar k_1|\to\infty$)
 be negative $\lambda_{j,max}^\sigma<0$.
 Under these conditions only
 eigenvalues $\lambda_1^\sigma(\bar{k}_1)$ have zeros.
 The possibility that these eigenvalues have only one zero
 can always be realized. This can be demonstrated for small
 enough values of $\b_{ij}$ (so called weak coupling
 approximation, see the next section) which in the limit
 $\b_{ij}=0$ for all $i,j$ gives a very simple behavior of
 the eigenvalues. For instance, for
  $\Delta_{j+1}-\Delta_j$ large enough
and
$\min\limits_j(\sqrt{\Delta_j}+\alpha_j)>
\max\limits_j(-\sqrt{\Delta_j}+\alpha_j)$
  the straight line
  $\lambda_1^\s(\bar k_1)$ never intersects with the
  hyperbolas $\lambda_j^\s(\bar k_1)$ so that small
  perturbations coming from small non-zero $\beta_{jl}$-values
  (in a physical terminology these perturbations shift the
  zero width resonances from the real energy axis
   to the complex plane)
  do not change the monotonous behavior of
  $\lambda_1^\s(\bar k_1)$ and, hence, do not bring
  additional roots to the equation
  $\lambda_1^\s(\bar k_1)=0$.

Thus, we see that the minimal value of virtual states
with the absence of bound states is equal to all possible
sign combinations of $\s$ which is
$n_{vmin}=
\sum\limits_\sigma 1 =2^{N-1}$.
 Hence, the maximal possible number of resonances
is obtained by subtracting this number from the total
number of solutions, i.e.
\begin{equation}\label{nr1}
2n_{r,max}= N2^{N-1}-2^{N-1}=(N-1)2^{N-1}\,.
\end{equation}

\section{Weak coupling approximation \label{sec:WC}}

For the number of channels $N>2$ there is no way to get
analytical solutions of system \rf{algeq}, \rf{algeqa} but
if the coupling parameters $\b_{ij}$ are small enough
assuming analyticity of the roots of the Jost-matrix determinant
as functions of $\b_{ij}$
 a
perturbation technique may be developed.
In this section we
demonstrate this possibility by obtaining first order
corrections to unperturbed values of the roots of the Jost-matrix
determinant corresponding to $\b_{ij}=0$.

For the zero coupling, matrix $U_0$
 becomes diagonal
$U_0=\mbox{diag}(\alpha_1,\alpha_2,\ldots,\alpha_N)$
and the system \eqref{jdz}, \eqref{thrE} reduces to
\begin{eqnarray}
(\alpha_1-ik_{0,1})(\alpha_2-ik_{0,2})\ldots(\alpha_N-ik_{0,N})  =  0\,,\label{eqfa}\\
k_{0,j}^2-k_{0,1}^2+\Delta_j  =  0\,,\qquad j=2,\ldots,N\,,
\label{eqf}
\end{eqnarray}
where the additional subscript $0$ corresponds to the
uncoupled case. Its solutions have the form
\begin{eqnarray}
 k_{0,1}^{(1,\sigma)} & = & -i\alpha_1\,,\quad k_{0,m}^{(1,\sigma)}  =  \sigma_m\sqrt{-\alpha_1^2-\Delta_m}\,, \quad m\neq 1\,, \nonumber \\
 k_{0,2}^{(2,\sigma)} & = & -i\alpha_2\,,\quad k_{0,m}^{(2,\sigma)}  = \sigma_m\sqrt{-\alpha_2^2+\Delta_2-\Delta_m}\,, \quad m \neq 2\,, \nonumber \\
 & \ldots & \nonumber  \\
 k_{0,N}^{(N,\sigma)} & = & -i\alpha_N\,,\quad k_{0,m}^{(N,\sigma)}  = \sigma_m \sqrt{-\alpha_N^2+\Delta_N-\Delta_m}\,, \quad m \neq N\,,
 \label{solf}
\end{eqnarray}
where $m=1,\ldots,N$. Let us explicitly indicate the meaning of sub- and superscripts in \eqref{solf}:
the second  subscript $m$ in
$k_{0,m}^{(j,\sigma)}$ corresponds to the channel, the first
superscript $j$  indicates a row number in \eqref{solf}
and $\sigma$ indicates one of all $2^{N-1}$ combinations of signs.
Thus, we see once again that the total number of
solutions of the system is $N2^{N-1}$ and it does not depend on
whether or not the coupling is absent. Note that every energy
level $E_j=-\alpha_j^2+\Delta_j$ corresponding to a row in
\eqref{solf} is $2^{N-1}$ fold degenerate. Below we show that
under a small coupling every degenerate level $E_j$ splits by
$2^{N-1}$ sub-levels and we will find approximate values of the
splitting. But the unperturbed $j$-th momentum corresponding to
this level
 simply equals $k_{0,j}^{(j,\sigma)}=-i\alpha_j$. Therefore,
 instead of our previous convention to express all
 quantities in terms of $k_1$, it is convenient here to express
 corrections to the $j$-th momentum produced by a perturbation in terms
 of unperturbed $j$-th momentum $\bar{k}_{0,j}^{(j,\sigma)}$. This is always possible due to the fact that
 all momenta have equal rights. But now we have to change our signs convention introduced in
 section \ref{sec:VS} where the first momentum $\bar{k}_1$ entered in the string
 $\s$ always with the positive sign ($\s_1=+$). Now we have $j$-th momentum $\bar{k}_j\in(-\infty,\infty)$ and
 $\sigma_j=+$ in string $\s$.

From \eqref{solf} we learn that no coupling implies no
finite-width resonances
but as we discuss below
the zeros lying above the first threshold
may be associated with zero-width resonances
which acquire a non-zero width under a small coupling.

From the first row of \eqref{solf} we conclude that the corresponding $2^{N-1}$ zeros with
$E_1=-\alpha_1^2$ are always below the first threshold (bound or virtual states).
Energy $E_n=-\alpha_n^2+\Delta_n$, $n=2,\ldots,N$,
may be positive with respect to the first
threshold and just these $(N-1)2^{N-1}$ zeros
are associated with the zero-width resonances.
According to our convention a resonance corresponds to a pair of
complex zeros.
Here we can easily compute the number of the zero-width
resonances, $n_{zwr}$, which is
 $n_{zwr}=(N-1)2^{N-2}$
which agrees with the
maximal number of possible resonances obtained in the previous
section.

Unperturbed $B^\s$ matrix we denote  $B^\s_0$
is diagonal
\begin{equation}
\label{vcm}
B_0^{\sigma}={\rm diag}
(\alpha_1+\sigma_1\bar{k}_1,\alpha_2+\sigma_2\bar{k}_2,\ldots,\alpha_N+\sigma_N\bar{k}_N)
\end{equation}
and its eigenvalues $\lambda_{0,j}^\sigma$ coincide with its diagonal entries
\begin{eqnarray}\label{laj}
\lambda_{0,j}^\sigma(\bar{k}_j)=\alpha_j+\bar{k}_j\,,\qquad
\lambda_{0,l}^\sigma(\bar{k}_j)=\alpha_l+\sigma_l\sqrt{\bar{k}_j^2+\Delta_l-\Delta_j}\,,\\
 l=1,\ldots, N\,,\qquad l\neq j\,.
\end{eqnarray}


For simplicity we assume all coupling parameters $\b_{ij}$
proportional to the same small parameter $\b$ so that
the perturbed $B^\s$ matrix reads
\be
B^{\sigma}=B_0^{\sigma}+\beta\mathcal{B}\,,\quad
 \mathcal{B}=||b_{jl}||\,,\quad b_{jj}=0\,,\quad
 j=1,\ldots,N\,.
 \ee
 Now
as it was mentioned above assuming analyticity of eigenvalues of this
matrix as functions of $\b$
 we can develop them in a Taylor series with respect to
 $\b$,
\begin{equation}\label{ev}
\tilde{\lambda}_j^\sigma=
\lambda_{0,j}^\sigma+\lambda_{1,j}^\sigma+\lambda_{2,j}^\sigma+
\ldots\,,
\end{equation}
where the first subscript number is just the power of $\b$.
First we notice that
the perturbation $\mathcal{B}$ has zero diagonal entries
which results in $\lambda^\s_{1,j}=0$.
To get the second order correction we are using the usual
Rayleigh-\Sc$\!\!$ perturbation approach which leads to
\begin{equation}\label{la2s}
\lambda_{2,j}^\sigma(\bar{k}_j)=
\beta^2\sum\limits_{l=1,\, l\neq j}^N\frac{b_{jl}^2}{\lambda_{0,j}^\sigma(\bar{k}_j)-
\lambda_{0,l}^\sigma(\bar{k}_j)}\,.
\end{equation}
In what follows we also assume that we can neglect the
higher-order corrections to the eigenvalues.

Actually, our aim is to find corrections
to the unperturbed degenerate $j$-th  Jost-matrix determinant zero
given in \rf{solf}.
Assuming a
 Taylor series expansions for this root
 over the small parameter $\b$ indicating it now explicitly
 \be\label{sk1}
\bar{k}_j=\bar{k}_{0,j}^{(j,\s)}+\beta c_1+\beta^2 c_2+\ldots \ee
we find coefficients $c_1$ and $c_2$ from the equation
\begin{equation}\label{eqla}
\tilde{\lambda}_j^\sigma(\bar{k}_j)=
\lambda^\s_{0,j}(\bar k_j)+\lambda^\s_{2,j}(\bar k_j)=0\,.
\end{equation}
For that we develop $\tilde{\lambda}_j^\sigma(\bar{k}_j)$ in a
Taylor series in $\b$ parameter considering its $\b$ dependence as
given through $\bar{k}_j$ and \rf{sk1}. The term \rf{la2s}
contains the factor $\b^2$, therefore in its denominator we simply
put $\bar{k}_{0,j}^{j,\s}$  instead of $\bar{k}_j$. The $\bar
k_j$-dependence of the term $\lambda^\s_{0,j}(\bar k_j)$ is given
by \rf{laj} and its $\b$-dependence is obtained via \rf{sk1}.
Thus, the left hand side of equation \rf{eqla} is presented as a
series over the powers of $\b$ where every coefficient should
vanish. This leads to $c_1=0$ and
\begin{equation}\label{c2}
c_2=\sum\limits_{l=1,\, l\neq j}^N
\frac{b_{jl}^2}{\alpha_l+\sigma_l\sqrt{\alpha_j^2+\Delta_l-\Delta_j}}\,.
\end{equation}

Finally up to the second order in $\b$ we obtain the roots
of system
 \eqref{solf}
 \be\fl
\begin{array}{lll}
 k_1^{(1,\sigma)} & = & -i\alpha_1+i\sum\limits_{l=2}^N \frac{\beta^2b_{1l}^2}{\alpha_l+\sigma_l\sqrt{\alpha_1^2+\Delta_l}}\,,\label{solpt}
 \quad k_m^{(1,\sigma)}  = \sigma_m \sqrt{\left(k_1^{(1,\sigma)}\right)^2-\Delta_m}\,,  \nonumber \\
  k_2^{(2,\sigma)} &  =  & -i\alpha_2+i\sum\limits_{l=1,\, l\neq 2}^N
 \frac{\beta^2b_{2l}^2}{\alpha_l+\sigma_l\sqrt{\alpha_2^2+\Delta_l-\Delta_2}}
 \,,
 \quad k_m^{(2,\sigma)}  = \sigma_m\sqrt{\left(k_2^{(2,\sigma)}\right)^2+\Delta_2-\Delta_m}\,,  \nonumber \\
 & \ldots &  \nonumber  \\
 k_N^{(N,\sigma)} & =  & -i\alpha_N+i\sum\limits_{l=1}^{N-1} \frac{\beta^2b_{Nl}^2}{\alpha_l+\sigma_l\sqrt{\alpha_N^2+\Delta_l-\Delta_N}}
 \,,\quad k_m^{(N,\sigma)}  =\sigma_m \sqrt{\left(k_N^{(N,\sigma)}\right)^2+\Delta_N-\Delta_m}\,.
\end{array}\label{kkk}
\ee Here each row is obtained by applying equations \eqref{la2s},
\eqref{sk1}, \eqref{eqla} and \eqref{c2} for $j=1,\ldots,N$,
respectively, and $m=1,\ldots,N$, $m\neq j$ for each $j$. The
square roots in the last column of \rf{kkk} should be expanded
in Taylor series up to $\beta^2$.

From here it is easily seen that, when
$\alpha_m^2<\Delta_m$,
purely imaginary unperturbed zeros $k_m=-i\alpha_m$
move from the axes to the complex plane due to the real part of corrections.
For instance for $k_2$, the real part reads
${\pm\beta^2\sqrt{\Delta_2-\alpha_2^2}/(\alpha_1^2-\alpha_2^2+\Delta_2)}$.
We thus confirmed the previous statement that zero width
resonances acquire non-zero widths.

\section{\label{sec:Cox}Zeros of the Jost-matrix determinant for $N=2$}

The particular case of two coupled channels is important both from practical and theoretical
point of view. Let us recall the following inequalities for the number of
the bound/virtual states and resonances obtained in sections \ref{sec:NBS}, \ref{sec:VS} and
\ref{sec:R}: $0\leq n_b\leq 2$, $0\leq n_r\leq 1$, $0\leq n_v\leq 4$.
The same inequalities are obtained for $N=2$ in \cite{pupasov:07,pupasov:08} from another approach.
The two-channel problem is the only one where one is able to get
analytic expressions for the Jost-determinant roots, i.e. to solve the direct
problem consisting in finding the positions of the bound/virtual states and resonances.
This possibility
is based on the fact that the roots of the algebraic equation of fourth, $(N2^{N-1})|_{N=2}=4$, order
may be expressed in radicals. Thus we obtain zeros as functions of parameters defining
the potential.
One may be interested in solving the inverse problem: to express
parameters of the potential from the knowledge about
positions of zeros of the Jost-matrix determinant. In principle, one may try to inverse
radicals, but we propose a more elegant way below.

To simplify the notations, we choose in this case $\Delta_2 \equiv
\Delta \neq 0$. The potential, which is known as the Cox potential
\cite{cox:64}, depends on three parameters appearing in matrix
\begin{equation}
U_0=  \left(
\begin{array}{cc}
\alpha_1 & \beta \\ \beta & \alpha_2
\end{array}
\right)
\end{equation}
and on the factorization energy $\cal E$ which is upper bounded.
The Jost-matrix determinant reads
\begin{equation}\label{det1}
f(k_1,k_2)\equiv\mbox{det}F(k_1,k_2)=
\frac{(k_1+i\a_1)(k_2+ia_2)+\b^2}{({k_1+i\kappa_1})(k_2+i\kappa_2)}.
\end{equation}


The system of equations \eqref{algeq}, \rf{algeqa} in this case
takes the simplest form
\begin{eqnarray}\label{sys}
&& k_1^2-k_2^2=\Delta, \\
&& (k_1+i\a_1)(k_2+i\a_2)+\b^2=0 \label{sys1}
\end{eqnarray}
and can be reduced to a fourth order algebraic equation with
respect to $k_1$ \be\label{k4}
k_1^4+ia_1k_1^3+a_2k_1^2+ia_3k_1+a_4=0\,. \ee
The coefficients $a_i$, $i=1,\ldots,4$ are given explicitly in \cite{pupasov:07}, (33a-d).
Momentum $k_2$ can
be found from \be\label{k5}
k_2(ik_1-\a_1)=\a_2(k_1+i\a_1)-i\b^2\,, \ee which is a direct
implication of \eqref{sys1}. Equations \eqref{k4} and \eqref{k5}
are particular case of the system \eqref{k1}, \eqref{k2}, \ldots,
\eqref{kN} for $N=2$ in accordance with our general discussion in
section~\ref{sec:Pr}.

Let us assume we have found two of the roots of system
\eqref{sys}, \eqref{sys1} we denote $(k_1^{(1)},k_2^{(1)})$ and
$(k_1^{(2)},k_2^{(2)})$, which clearly are functions of parameters
$\a_1$ and $\a_2$. Their dependence on parameters $\b$ and $\Delta$ is not
important for the moment, since both $\b$ and $\Delta$ assumed to be fixed. Being put back to \eqref{sys1} the
equation reduces twice to identity for any values of $\a_1$ and
$\a_2$, which we write as
\begin{eqnarray}
\label{sys11a}
& (k_1^{(1)}+i\a_1)(k_2^{(1)}+i\a_2)+\b^2=0\,,\\
& (k_1^{(2)}+i\a_1)(k_2^{(2)}+i\a_2)+\b^2=0\,.\label{sys11b}
\end{eqnarray}
The reason why we replaced the identity sign by the equality sign
is that these equations may be considered as an implicitly written
inverted dependence of $\a_{1,2}$ on the set of parameters
$k_{1,2}^{(1,2)}$. We  may thus fix arbitrary values for
$k_{1,2}^{(1,2)}$ and find from \eqref{sys11a}, \eqref{sys11b}
$\a_1$ and $\a_2$ in terms of
 $k_{1,2}^{(1,2)}$ which is a much easier task than
 finding an explicit dependence of $k_{1,2}^{(1,2)}$ on
 $\a_1$ and $\a_2$.
For that
 we have to solve, e.g. for
 $\a_1$, the following second order equation
\be
\a_1^2-\a_1i(k_1^{(1)}+k_1^{(2)})-k_1^{(1)}k_1^{(2)}+\b^2\frac{R_1}{R_2}=0\,,
\ee with
$R_1 = k_1^{(2)}-k_1^{(1)}$ and $R_2 = k_2^{(2)}-k_2^{(1)}$ which
easily follows from
\eqref{sys11a} and \eqref{sys11b}. From here we find
\begin{eqnarray}\label{alp1}
\a_1 &=& \frac{1}{2}\left[i(k_1^{(1)}+k_1^{(2)})
\pm\sqrt{-R_1^2-4\b^2R_1/R_2}\right],\\
\label{alp2} \a_2 &=&
\frac{1}{2}\left[i(k_2^{(1)}+k_2^{(2)})\mp\sqrt{-R_2^2-4\b^2R_2/R_1}\right].
\end{eqnarray}
The upper (resp., lower) sign in \eqref{alp2} corresponds to the
upper (resp., lower) sign in \eqref{alp1}. The values of
$k_1^{(1,2)}$ and $k_2^{(1,2)}$ should be chosen so as to warranty
the reality of parameters $\a_{1,2}$.

Once two roots are fixed, \eqref{k4} reduces to a second-order
algebraic equation $\mathcal{Q}_2(k_1)=0$ for the two other roots
$k_1^{(3)}$ and $k_1^{(4)}$ thus providing an implicit but rather
simple mapping between the roots of system \eqref{sys},
\eqref{sys1} and the set of parameters $(\a_1,\a_2,\b)$.
Polynomial $\mathcal{Q}_2(k_1)$ is the ratio of the polynomial
appearing in \eqref{k4} and
${\mathcal{P}_2(k_1)=k_1^2-k_1(k_1^{(2)}+k_1^{(1)})+k_1^{(2)}k_1^{(1)}}$,
i.e.,
\[
k_1^4+ia_1k_1^3+a_2k_1^2+ia_3k_1+a_4=\mathcal{P}_2(k_1)\mathcal{Q}_2(k_1)\,.
\]
From here we find, with the explicit expression for coefficients $a_i$, $i=1,\ldots,4$ \cite{pupasov:07},
\bea\nn
\mathcal{Q}_2(k_1)=(k_1+i\a_1)^2+k_1(k_1^{(2)}+k_1^{(1)})+\\
(2i\a_1+k_1^{(2)}+k_1^{(1)})(k_1^{(2)}+k_1^{(1)})+\a_2^2-\Delta-k_1^{(1)}k_1^{(2)}
\nn \eea
and, hence,
%
\bea\label{k34a}
k_1^{(3)}=\frac{1}{2}\left[\mp i\sqrt{-R_1^2-4\b^2R_1/R_2}+\sqrt{D_1}\right], \\
k_1^{(4)}=\frac{1}{2}\left[\mp
i\sqrt{-R_1^2-4\b^2R_1/R_2}-\sqrt{D_1}\right],\label{k34b} \eea
where $D_1=R_1^2+4\b^2\frac{R_2}{R_1}+4k_1^{(2)}k_1^{(1)}$. The
sign before the first square root in \eqref{k34a} and
 \eqref{k34b} should be
chosen in accordance with the signs in \eqref{alp1} and
\eqref{alp2}.

To find $k_2^{(3,4)}$ we do not need to solve any equation. We
simply notice that the equation $\mbox{det}F(k_1, k_2)=0$
is invariant under the transformation $k_1 \leftrightarrow k_2$,
$\a_1 \leftrightarrow \a_2$, $\Delta \leftrightarrow -\Delta$.
This means that being transformed according to these rules
equations \eqref{k34a} and \eqref{k34b}
 give us the $k_2$ values:
\bea \label{p34} k_2^{(3)}= \frac{1}{2} \left[\mp i
\sqrt{-R_2^2-4\b^2R_2/R_1}-\sqrt{D_2}\right], \\
k_2^{(4)} = \frac{1}{2} \left[\mp i
\sqrt{-R_2^2-4\b^2R_2/R_1}+\sqrt{D_2}\right], \eea
where $D_2 = R_2^2+4\b^2\frac{R_1}{R_2}+4k_2^{(2)}k_2^{(1)}$.

\begin{table}\caption{\label{tab2}
Possible mappings between some experimental data and the Cox
potential parameters.}
\begin{center}
\begin{tabular}{|l||l||l||l|}
\hline Experimental & Fixed  & Free  & Restrictions  \\
data & parameters  & parameters &
\\
 \hline
 $\Delta\,, E_r\,, E_i$ & $\a_1\,, \a_2\, $   &
 $\kappa_1, \b$ & $\b\geq \sqrt{-k_rp_r}$
\\ \hline
$\Delta\,, E_b=-\lambda_b^2\,, E_r\,, E_i $ &$\a_1\,, \a_2\,, \b$
 & $\kappa_1$ & $\kappa_1>\lambda_b$
\\ \hline
 $\Delta\,, E_{1,2}=-\lambda_{1,2}^2$ & $\a_1\,, \a_2\, $   &
 $\kappa_1, \b$ & $\kappa_1>\lambda_2>\lambda_1$
\\ \hline
 $\Delta\,, E_{b}=-\lambda_b^2 $ &  $\a_2\, $ & $\kappa_1\,, \b\,, \alpha_1$ &
 $\kappa_1>\lambda_b$
\\ \hline
\end{tabular}
\end{center}
\end{table}

Two initial zeros $(k_1^{(1)},k_2^{(1)})$, $(k_1^{(2)},k_2^{(2)})$
and threshold difference $\Delta$ are assumed to be known from
the experiment. For instance, these zeros may correspond to a visible
Feshbach resonance or two bound states.
The possible cases for initial data are summarized in Table~\ref{tab2}.
The first row of Table \ref{tab2} corresponds to the case where
the position of the resonance is known (see section~\ref{sec:1r} below).
The second row corresponds to the case where the positions  of both the
resonance and one bound state are known, which allows one to fix
maximal number of parameters. The third row corresponds to the case
where the positions  of two bound states are known
(see section~\ref{sec:2b} below). The last row corresponds
to the special case when only one zero may be fixed from
experimental data. The free parameters in Table \ref{tab2} allow either for isospectral
deformations of the potential or for fits of additional
experimental data as, e.g., scattering lengths (see e.g. \cite{pupasov:07,pupasov:08}).
The restriction
on the factorization energy is deduced from the regularity condition of the potential \eqref{posU}.
The restriction on the coupling parameter $\beta$ is
explained below (see \eqref{ren} in section~\ref{sec:1r}).
Let us now consider examples corresponding to the first three rows of Table \ref{tab2}.

\subsection{One resonance.\label{sec:1r}}

A resonance corresponds to a pair of complex
 roots $k_1^{(1)}$ and $k_1^{(2)}$ of the Jost-matrix determinant
 such that  $ik_1^{(1)}$ and $ik_1^{(2)}$ are mutually
 complex conjugate.
Therefore we assume equations \eqref{sys} and \eqref{sys1} to have
two complex roots. Let us define their first-channel components as
 \begin{equation}
\label{k12} k_1^{(1)} = k_r+ik_i\,,\quad k_1^{(2)} =
-k_r+ik_i\,,\quad k_{i}\in\Bbb R\,,\quad k_{r}\in\Bbb R\,, \quad k_r>0\,,
 \end{equation}
and write the corresponding energies, $\left(k_1^{(1,2)}\right)^2$, as $E_r \pm i
E_i$, where we also assume $E_i>0$ (which means that the upper
sign corresponds to $k_1^{(1)}$ or $k_1^{(2)}$, depending on the
sign of $k_i$). We would like to choose as parameters the
threshold difference $\Delta$, as well as the real and imaginary
parts of the resonance complex energy, $E_r, E_i$. As exemplified
below, these can correspond to physical parameters of a visible
resonance in some (but not all) cases. In terms of these
parameters, $k_r$ and $k_i$ are expressed as
 \begin{equation}
\label{kroots} k_r= \frac{E_i}{\sqrt
2}\,\left[\sqrt{E_r^2+E_i^2}-E_r\right]^{-1/2}, \quad k_i=\pm
\frac{1}{\sqrt 2}\,\left[\sqrt{E_r^2+E_i^2}-E_r\right]^{1/2}.
\end{equation}
In the second channel the roots
\[
k_2^{(1)} = p_r+ip_i\,,\quad k_2^{(2)} = -p_r+ip_i\,,
\]
can be found from the threshold condition yielding
\bea\label{proots} p_r&=& -\frac{1}{\sqrt
2}\,\left[\sqrt{E_i^2+(E_r-\Delta)^2}+E_r-\Delta\right]^{1/2}\!\!, \\
 p_i&=& \mp\frac{E_i}{\sqrt2}\,\left[\sqrt{E_i^2+(E_r-\Delta)^2}+
 E_r-\Delta\right]^{-1/2}\!\!\!\!\!\!.
\label{proots2} \eea
The upper (resp., lower) sign in \eqref{kroots} corresponds to the
upper (resp., lower) sign in \eqref{proots2}, which means that, for
a given zero, the signs of $k_i$ and $p_i$ are
opposite. Moreover, equations \eqref{kroots} and~\eqref{proots}
show that, for a given zero, the signs of $k_r$ and $p_r$ are also
opposite. This implies that, for the Cox potential, the complex
resonance zeros (or scattering-matrix poles) are always in
opposite quadrants in the complex $k_1$ and $k_2$ planes.
\begin{figure}
\begin{center}
\epsfig{file=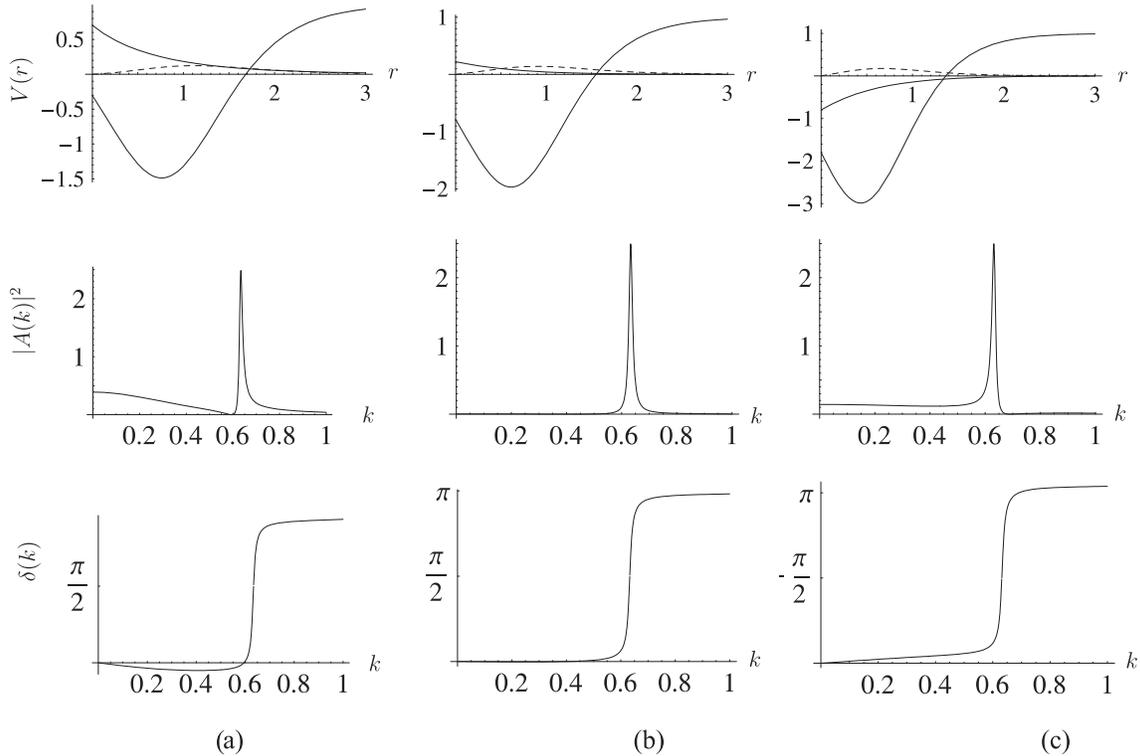, width=15cm} \caption{\small The Cox
potential without bound state and with one visible resonance of
energy $E_r=0.4$ and width
$\Gamma=0.02$, for $\Delta=1$ and $\beta=0.1$ (first row, solid
lines for $V_{11}$ and $V_{22}+\Delta$, dashed
line for $V_{12}$), with the corresponding partial cross
section (second row) and phase shifts (third row) for (a)
$\kappa_1 = 0.5$; (b) $\kappa_1=0.7$; (c) $\kappa_1 = 1$.
\label{figExR}}
\end{center}
\end{figure}
This has important consequences for physical applications: for a
resonance to be visible, one of the corresponding zero has to lie
close to the physical positive-energy region, i.e., close to the
real positive $k_1$ axis and close to the region made of the real
positive $k_2$ axis and of the positive imaginary $k_2$ interval:
$[0,i\sqrt{\Delta}]$. Consequently, the only possibility for a
visible resonance with the Cox potential is that of a Feshbach
resonance, only visible in the channel with lowest threshold, with
an energy lying below threshold $\Delta$. At higher resonance
energies, the corresponding zero is either close to the
$k_1$-plane physical region (and far from the $k_2$-plane one) or
close to the $k_2$-plane physical region (and far from the
$k_1$-plane one); it cannot be close to both physical regions at
the same time, hence it cannot have a visible impact on the
coupled scattering matrix. Here, we illustrate the case of a
visible resonance, which is the most interesting from the physical
point of view. It corresponds to the lower signs in \eqref{kroots}
and~\eqref{proots2}, with a resonance energy $E_r$ such that
$0<E_r<\Delta$, and a resonance width $\Gamma=2 E_i$ such that
$E_i<E_r$.

Note, that for non-zero values of the parameters $k_r$ and $p_r$
(which have opposite signs), the coupling parameter $\b$ cannot be
infinitesimal: because $\a_{1}$ and $\a_{2}$ have to be real, $\b$
is restricted to satisfy the inequality
\begin{equation}\label{ren}
\b \ge \sqrt{-k_r p_r}\,.
\end{equation}

To get a potential with one bound state at energy $-\lambda_b^2$,
we choose the lower signs in \eqref{alp1}, \eqref{alp2}. We then
get for $k_1^{(3)}(\beta)$ an expression similar to \eqref{k34a},
\eqref{k34b}, from which the value of $\beta$ can be found by
solving the bi-squared equation $ k_1^{(3)}(\b)=i \lambda_b$.

Let us now choose explicit parameters. First, we put $\Delta=1$.
To get a visible resonance, we put $E_r=0.4$, $E_i=0.01$ (which
corresponds to a resonance width $\Gamma=0.02$), and $\b=0.1$.
Using \eqref{alp1},~\eqref{k12} and~\eqref{kroots}, one finds
$\a_1=0.76938$ and $\a_2=-0.766853$ (we choose the upper signs
\eqref{alp1}, \eqref{alp2}). The factorization energy, $\cal E$,
is not constrained in this case: it just has to be negative. The
Cox potential with one resonance and two virtual states
$E_{v1}=-0.560473$, $E_{v2}=-0.599544$
is shown in the first row of figure~\ref{figExR}.

The diagonal elements of the potentials, $V_{11}$ and $V_{22}+\Delta$,
are plotted with solid lines,
while $V_{12}$ is plotted with dashed lines.
Parameter $\kappa_1$ is responsible for the isospectral deformation of the potential
which results in the behavior of the phase shifts.
The second row of figure~\ref{figExR} shows the corresponding partial cross sections,
where the resonance behavior is clearly seen, as well as the evolution of the low-energy cross section,
which is related to the scattering length.
The last row of figure~\ref{figExR} shows the corresponding phase shifts for the open channel,
where a typical Breit-Wigner behavior (see e.g.\ Ref.~\cite{taylor:72})
is seen for the resonance, as well as the evolution of the zero-energy phase-shift slope,
which is also related to the scattering length.

\subsection{Two bound states\label{sec:2b}}
\begin{figure}
\begin{center}
\begin{minipage}{12cm}
\epsfig{file=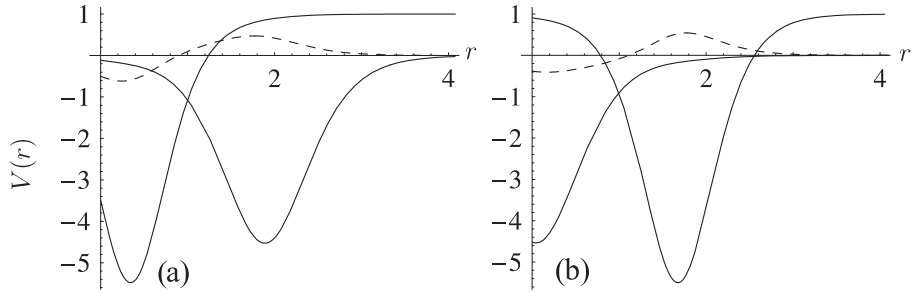, width=12cm} \caption{\small The Cox
potential (solid lines for $V_{11}$ and
$V_{22}+\Delta$, dashed line for $V_{12}$) with
two bound states at energies $E_1=-0.01$ and $E_2=-2.25$, for
$\Delta=1$, $\beta=0.1$ and $\kappa_1 = 1.51$. The left (resp.,
right) graphic corresponds to the upper (resp., lower) signs
 in \eqref{alp1} and \eqref{alp2}.  \label{figExB2}}
\end{minipage}
\end{center}
\end{figure}
Let us now construct a Cox potential with two bound states, and
hence no resonance \cite{pupasov:07}. We choose $k_1^{(1)}=0.1i$ and
$k_1^{(2)}=1.5i$ for these bound states and, as in the previous
example, we put $\Delta=1$ and $\b=0.1$. We thus have
$k_2^{(1)}=\sqrt{1.01} i$ and $k_2^{(2)}=\sqrt{3.25} i$, which
defines $R_2$ in \eqref{alp1}, \eqref{alp2}. Choosing the upper
signs in these equations, we find $\a_1=-0.112649$ and
$\a_2=-1.79557$, while for the lower signs, we get $\a_1=-1.48735$
and $\a_2=-1.0122$. The corresponding Cox potentials are shown in
figure~\ref{figExB2}.

\section{Conclusion}

A careful study of spectral properties of non-conservative
multichannel SUSY partners of the zero potential is given.
Our treatment is based on the analysis of the
Jost-matrix determinant zeros. Generalizing our previous
results for the two-channel case \cite{pupasov:07,pupasov:08},
we have shown that the zeros of the Jost-matrix determinant are the roots
of an $N2^{N-1}$th-order
algebraic equation.
The number of bound states $n_b$
is restricted by the number of channels, $0\leq n_b \leq N$. The upper
bound for the number of resonances is $(N-1)2^{N-2}$. The generalization
is based on the analysis of the behavior of the Jost-matrix eigenvalues.

In general, an algebraic equation of an order higher
than $4$ has no solutions in radicals. As a consequence, there are no
exact analytic solutions of
a spectral problem for a non-conservatively SUSY
transformed Hamiltonian with $N>2$.
Therefore, the problem of finding the approximate solutions appears to be actual.
Based on the usual Rayleigh-\Sc$\!\!$ perturbation theory
for the eigenvalues of the Jost matrix we develop an approximate
method for finding the zeros of the Jost-matrix determinant
in the case of a weak coupling between
channels.

An analytical study of the Jost-determinant zeros is carried out for the two-channel case
which implies an algebraic equation of the fourth order.
A suitable factorization of the fourth-order polynomial allows us to develop
a procedure which solves the inverse
spectral problem for this case.
  The effectiveness of the procedure
is illustrated by two
examples: a potential with one resonance and
a potential with two bound states.

\ack
We thank Daniel Baye for very useful discussions at several stages of this work.
AP is supported by Russian "Dynasty" foundation.
BFS is partially supported by grant RFBR-06-02-16719.
BFS thanks the National Fund for Scientific Research, Belgium,
for support during his stay in Brussels.
This text presents research results of the Belgian program P6/23 on interuniversity
attraction poles of the Belgian Federal Science Policy Office (BriX, Belgian Research Initiative on eXotic nuclei).

\end{document}